\documentclass{jpsj3}
\usepackage{txfonts}
\usepackage[usenames]{color}

\title{Decay of Resonance Structure and Trapping Effect in Potential
Scattering Problem of Self-Focusing Wave Packet}

\author{
Hironobu F{\sc ujishima}\thanks{E-mail address:
fujishima.hironobu@canon.co.jp}, Makoto M{\sc ine}$^1$\thanks{E-mail address:
mine@waseda.jp}, Masahiko O{\sc kumura}$^{2,3}$\thanks{E-mail address:
okumura.masahiko@jaea.go.jp} and Tetsu Y{\sc ajima}$^4$\thanks{E-mail address:
yajimat@is.utsunomiya-u.ac.jp}
}
\inst{
Optics R\&D Center, CANON INC., 23-10 Kiyohara-Kogyodanchi,
Utsunomiya 323-3298, Japan \\
$^1$Waseda University Honjo Senior High School, 1136 Nishitomida Honjo
Saitama 367-0035 Japan \\
$^2$CCSE, Japan Atomic Energy Agency, 5-1-5 Kashiwanoha Kashiwa Chiba 277-8587 Japan\\ 
$^3$CREST (JST), Kawaguchi, Saitama 332-0012, Japan \\
$^4$Department of Information Systems Science, Graduate School of
Engineering, Utsunomiya University, Yoto 7-1-2, Utsunomiya 321-8585,
Japan 
}
\abst{
Potential scattering problems governed by the time-dependent
Gross-Pitaevskii equation are investigated numerically for various
values of coupling constants. The initial condition is assumed to have
the Gaussian-type envelope, which differs from the soliton
solution. The potential is chosen to be a box or well type. We estimate
the dependences of reflectance and transmittance on the width of the
potential and compare these results with those given by the stationary
Schr\"odinger equation. We attribute the behaviors of these quantities
to the limitation on the width of the nonlinear wave packet. The coupling
constant and the width of the potential play an important role in
the distribution of the waves appearing in the final state of scattering.
}
\kword{Gross-Pitaevskii equation, nonlinear Schr\"odinger equation,
potential scattering, Gaussian initial condition, numerical analysis,
Bose-Einstein condensation} 
\begin{document}
\maketitle

\section{Introduction}
The nonlinear Schr\"odinger equation (NLSE), which has a cubic
nonlinear term,
\begin{equation}
\mathrm{i}\phi_{t}+\phi_{xx}+2|\phi|^2\phi=0\label{NLSE},
\end{equation}
appears in various fields of physics\cite{Karpman}. The NLSE can be
derived as an amplitude equation of a system whose dispersion
relation depends dominantly on the square of the amplitude. Among such systems
is the envelope motion of coupled nonlinear oscillators with cubic
interaction\cite{Newell}. In nonlinear optics, both the self-focusing
effect in two-dimensional (2D) systems and optical soliton propagation
in one-dimensional (1D) systems are governed by the
NLSE\cite{Hasegawa,Agrawal}. Another important example is the
Bose-Einstein condensed (BEC) system\cite{Dalfovo,Pitaevskii} in which the macroscopic wave function of condensate atoms appears as the order
parameter accompanied with the spontaneous breakdown of the $U(1)$
gauge symmetry\cite{Okumura}. In this case, the NLSE is regarded as the mean-field approximation of the Heisenberg equation for field
operators and describes the time evolution of this macroscopic wave
function with good accuracy\cite{Dalfovo,Pitaevskii}.  

One of the striking features of 1D NLSE is its integrability. In
particular, exact solutions under a given initial condition can be
uniquely solved by the inverse scattering transformation (IST)
method\cite{Gardiner,Zakharov}. For the sech-type initial condition with
suitable amplitude, the entire initial value problem is solved analytically and
results in the {\it N}-soliton solution\cite{Satsuma}. However, the time evolution of the wave packet from the nonsoliton initial condition is relatively unclear since analytic expressions are rarely feasible. 

On the other hand, in the BEC system, it is natural to assume the
existence of an external field to express the effect of gravity or
quadratic traps for atoms\cite{Salasnich}. Thus, the term of external
field is added to the conventional NLSE (\ref{NLSE}), and the equation is
called the time-dependent Gross-Pitaevskii equation (TDGPE). So far,
most analytic studies have assumed only linear or quadratic potentials,
 in which case the integrability of the systems are not spoiled
and one can obtain analytical results by systematic application of the
IST method\cite{Chen,Balakrishnan}. In such cases, soliton initial
conditions are also assumed. 

However, we can also consider spatially localized potentials where the
integrability is manifestly violated. It is important to evaluate the
role of nonlinearity on such potential scattering problems as the
tunneling effect. In particular, time-dependent analysis of moving
wave packets is intriguing since \textit{in situ} observation of
condensed atoms is possible in the BEC system\cite{Dalfovo,Pitaevskii}, although most studies deal with these kinds of problems as stationary ones\cite{Hyouguchi}. 

When we analyze the stationary potential scattering problems, we completely adopt the wavelike nature and the resonant phenomena brought about by the interference effect. On the the other hand, the spatially localized pulse is expected
to exhibit the so-called wave packet effect in the scattering
process. In addition, nonlinear effects are also interesting in the potential scattering problem. 

To clarify the definition, we shall use ``the wave packet effect" and ``the nonlinear effect" in the following senses throughout this paper. a) The wave packet effect means a phenomenon caused by the fact the wave packet is localized as a result of superposing many monochromatic waves. b) The nonlinear effect means a phenomenon that cannot be interpreted without considering the nonlinearity.   
To investigate the influence of these effects on the potential scattering problem, it is necessary to trace the dynamics of the system. Some authors have reported on this kind of problem assuming soliton initial conditions\cite{Frauenkron,Sakaguchi}. However, examples that take nonsoliton solutions as initial conditions have been rare because of extra complexities. In this paper, we numerically trace and examine the dynamics of the wave packets governed by 1D TDGPE with the box- or well-type potential under the Gaussian-type initial conditions. 

This paper is organized as follows. In the next section, the
nonsoliton dynamics of the wave packets without external field are
analyzed. In \S3, we evaluate and characterize the nonlinear
wave packet in terms of the reflectance or transmittance, changing the
magnitude of the nonlinearity, position of the initial wave packet,
and the width of the potential. The section \S4 is devoted to the
discussion, and we interpret the results on the basis of wave packet and nonlinear effects. Extra complexities intrinsic to nonsoliton
initial conditions are also argued in detail. The summary is given in
the \S5. 

\section{Time Dependent Gross-Pitaevskii Equation and Scattering Problem}
In this section, we briefly summarize mathematical descriptions of
the system to be considered. We restrict ourselves to the 1D case
throughout this paper. By virtue of scale transformation, we can put
both coefficients $\phi_{t}$ and $\phi_{xx}$ of the TDGPE to
be unity and we shall consider 
\begin{equation}
\mathrm{i}\phi_{t}=-\phi_{xx}+V(x)\phi+g|\phi|^2\phi\label{TDGPE},
\end{equation}
where $V(x)$ is an external potential applied to the system, and $g$
is the coupling constant. To investigate the wave packet and nonlinear effects, the sign of $g$ is important and we discard the possibility of $g>0$ throughout this paper. If we take $g$ to be positive, which means a repulsively interacting field, the wave packet immediately expands and this rapid diffusing makes the amplitude of the wave packet very small. Therefore, the excitation of higher harmonic waves is extremely suppressed, and less manifestation of the nonlinear effect is expected. Moreover, these widespread wave packets share most of the scattering features with stationary plane waves in the linear limit. Therefore, we focus on the $g<0$ case, which means an attractively interacting field, in order to investigate the wave packet and nonlinear effects on the potential scattering problem. According to the theory of the partial differential equation, a finite and unique solution of eq.~(\ref{TDGPE}) exists for arbitrary initial conditions for the 1D case, and instability or explosion of the solution observed in multidimensions never occurs even if we take $g$ to be negative.

Hereafter, we shall normalize the wave function $\phi$ as
$\int_{\Bbb{R}}|\phi|^2\mathrm{d}x=1$. The initial condition of the
wave packet is fixed to be the Gaussian type, 
\begin{equation}
\phi(x,0) = \frac{1}{\sqrt[4]{\pi}}
 \mathrm{e}^{-\frac{1}{2}(x+x_{0})^2+\mathrm{i}v(x+x_{0})}
 \label{Gaussian}, 
\end{equation}
where $-x_{0}$ is the position of the center of the initial wave
packet and $v$ gives half of its velocity. We assume right-forward
propagation of the wave packet, i.e., $x_{0}>0$ and $v>0$. The energy
functional $E$, the Hamiltonian of the system, is defined as
\begin{equation}
E = \int_{\Bbb{R}}\left(|\phi_{x}|^2 + V(x)|\phi|^2 +
 \frac{1}{2}g|\phi|^4\right)\mathrm{d}x \label{energy}.  
\end{equation}
Equation (\ref{TDGPE}) can be derived straightforwardly from the
Hamiltonian through standard canonical procedure. Since we have
assumed $g$ to be negative, $E$ might take a negative value. In fact,
for the initial wave packet located sufficiently far from the
potential, the initial value of $E$ becomes negative under the
condition 
\begin{equation}
\frac{1}{2}+v^2+\frac{g}{\sqrt{8\pi}}<0\label{condition}.
\end{equation}
We consider the box- and well-type external potentials,
\begin{align}
V_{\mathrm{box}} & = \theta(x)-\theta(x-a)\label{box}, \\
V_{\mathrm{well}} & = -V_{0}(\theta(x)-\theta(x-a))\label{well},
\end{align}
where $a$ is the width of the potential and $\theta(x)$ denotes the
step function. We define the reflectance and transmittance from these
potentials as 
\begin{align}
R_{\mathrm{box}} & = \lim_{t\rightarrow\infty}
 \int_{-\infty}^{0}|\phi|^2 \mathrm{d}x\label{rbox}, \\
R_{\mathrm{well}} & = \lim_{t\rightarrow\infty}
 \int_{-\infty}^{-b}|\phi|^2\mathrm{d}x\label{rwell}, \\
T_{\mathrm{well}} & = \lim_{t\rightarrow\infty}
 \int_{a+b}^{\infty}|\phi|^2\mathrm{d}x\label{twell}. 
\end{align}
The reason why we introduce $b$ for the definitions of $R_{\rm
well}$ un eq.~ (\ref{rwell}) and $T_{\rm well}$ unh eq.~(\ref{twell}) is as follow. 
For the well-type potential, part of the wave packet is trapped by the
potential well and oscillates around the potential area. The distance
of $b$ is provided as a margin to distinguish the trapped portion and
reflected or transmitted ones. The trapped portion never completely
separates from the other parts of the wave packet, and  continues to
exchange a very small amount of their norms, and the limits in
eqs. (\ref{rwell}) and (\ref{twell}) do not exist in the strict
meaning. However, for an evaluating measure, we use the values at
$t=80$ as if they were limiting ones. In the next sections, we vary
$g$, $x_{0}$, and $a$ and investigate their influence on 
$R_{\mathrm{box}}$ and $T_{\mathrm{well}}$. For numerical integration, 
we employ the symplectic Fourier method\cite{Sasa} throughout this paper.

\section{Results}
In this section, the effects of nonlinearity on free propagation and
potential scattering problems under the Gaussian initial condition
are considered.

\subsection{Free propagation}
\begin{figure}[h]
\begin{center}
\includegraphics{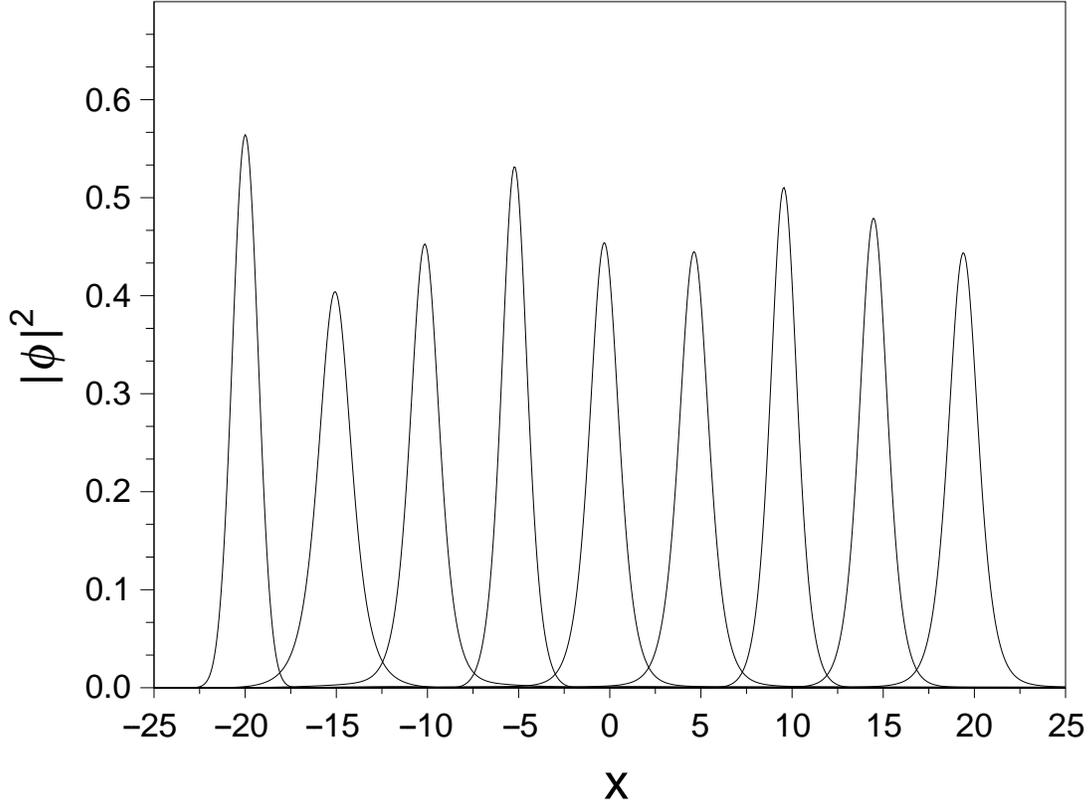}
\end{center}
\caption{Free propagating breather-like motion of a wave packet
 starting from the initial wave packet given in eq.~(\ref{Gaussian}) with
 $x_{0}=20$, $v=\sqrt{1.5}$, and $g=-4$. The nine wave packets show
 $|\phi|^2$ at $t$=0, 2, \ldots, 14, and 16 from the left to the
 right.} \label{f1}
\end{figure}

\begin{figure}
\begin{center}
\includegraphics{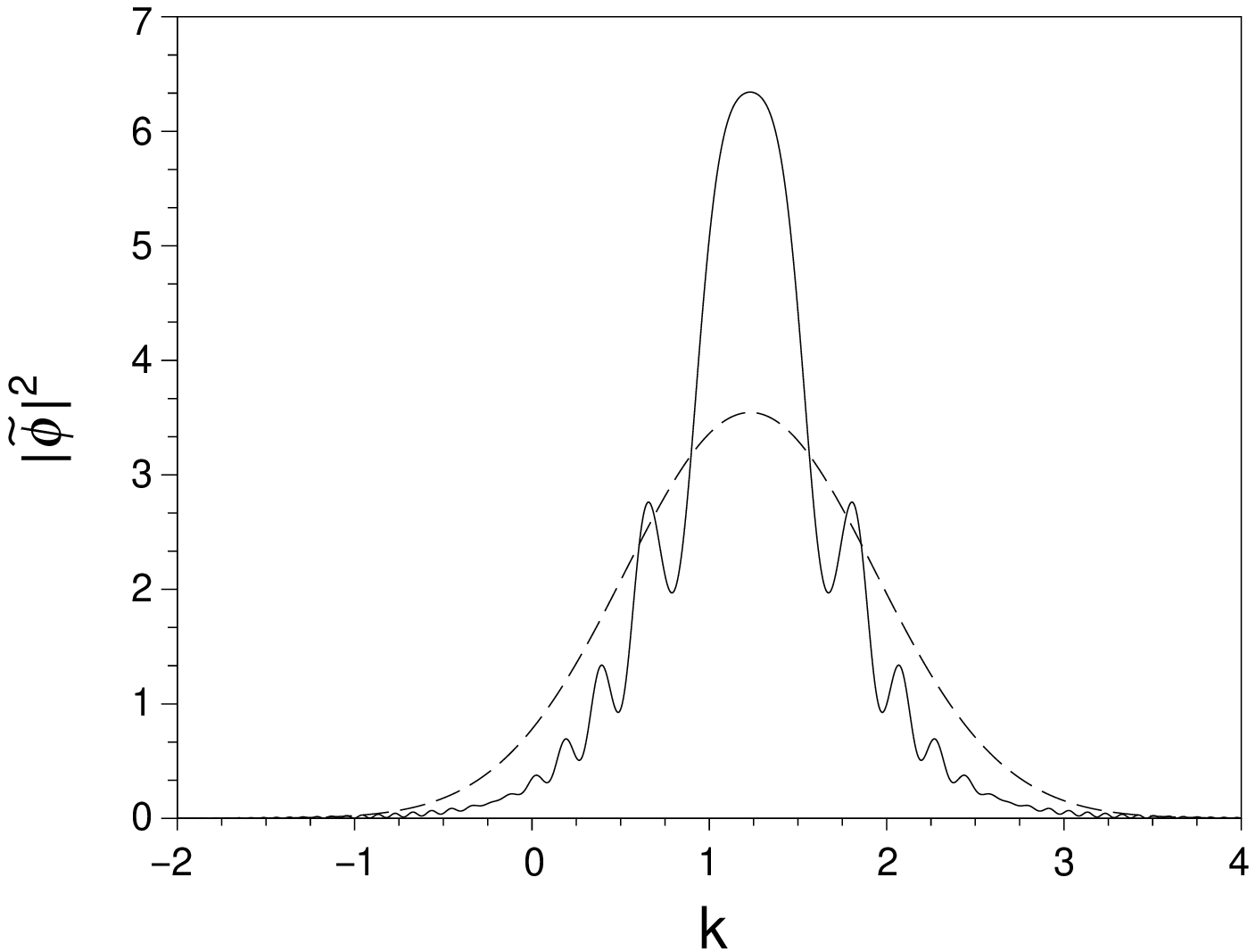}
\end{center}
\caption{Solid line shows $|\tilde{\phi}|^2$, the freely propagating breather-like wave packet observed in the wave-number space at $t=16$. The parameters are the same as the ones used in the last of Fig.~\ref{f1}. The dashed line is for $t=0$.} 
\label{f2}
\end{figure}

In this subsection, we discuss the free propagation of a wave packet where
no external potential exists. In this case, the 1-soliton solution of
eq.~(\ref{TDGPE}) with $V(x) = 0$ can be written as 
\begin{equation}
\phi(x,t) = \sqrt{\frac{2}{-g}} \eta\,\mathrm{sech}(x\eta+2t\eta\xi)
 \mathrm{e}^{\mathrm{i}\{x\xi-t(\eta^2-\xi^2)\}}, 
\end{equation}  
where $\eta$ and $\xi$ are independent parameters and control the amplitude and velocity of the soliton, respectively. Once this form of
solution is taken to be the initial condition, it never diffuses and
mantains its own shape during time evolution. 

However, the Gaussian initial condition (\ref{Gaussian}) leads to
breather-like solutions. We show the time evolution of the wave
profile $|\phi|^2$ in Fig.~\ref{f1}. We also calculate the wave function in wave-number space $\tilde{\phi}$ as
\begin{equation}
\tilde{\phi}(k,t)=\frac{1}{\sqrt{2\pi}}\int_{-\infty}^{\infty}\phi(x,t)e^{\mathrm{i}kx}\mathrm{d}x.
\end{equation}
We show a snapshot of $|\tilde{\phi}|^2$ taken at $t=16$ in Fig.~\ref{f2}.
 In the wave-number space, the breathing motion is also
observed and a notched structure grows on the surface of the wave
packets. This structure is a result of repeated expansion and
contraction in the wave-number space, i.e, expansion by the higher
harmonic excitation and contraction by the dispersion effect
(suppression of higher harmonic excitation). 

The breathing motion is a genuine nonlinear effect, since we would have observed that $|\phi|^2$ is simply diffusing and $|\tilde{\phi}|^2$ remains still if $g=0$. It is known that any solitary waves governed by eq.~(\ref{TDGPE}) with $V(x)=0$ finally split into a complete soliton part and a small oscillating tail (radiation) which rapidly leaves the soliton part in the long run\cite{Ablowitz}. Therefore, this breather-like behavior is considered to have a finite lifetime and the decay process is a rather transient phenomenon. Fortunately, this lifetime is sufficiently long to observe the breather-like motion.

\subsection{Box-type potential}
\begin{figure}
\begin{center}
\includegraphics{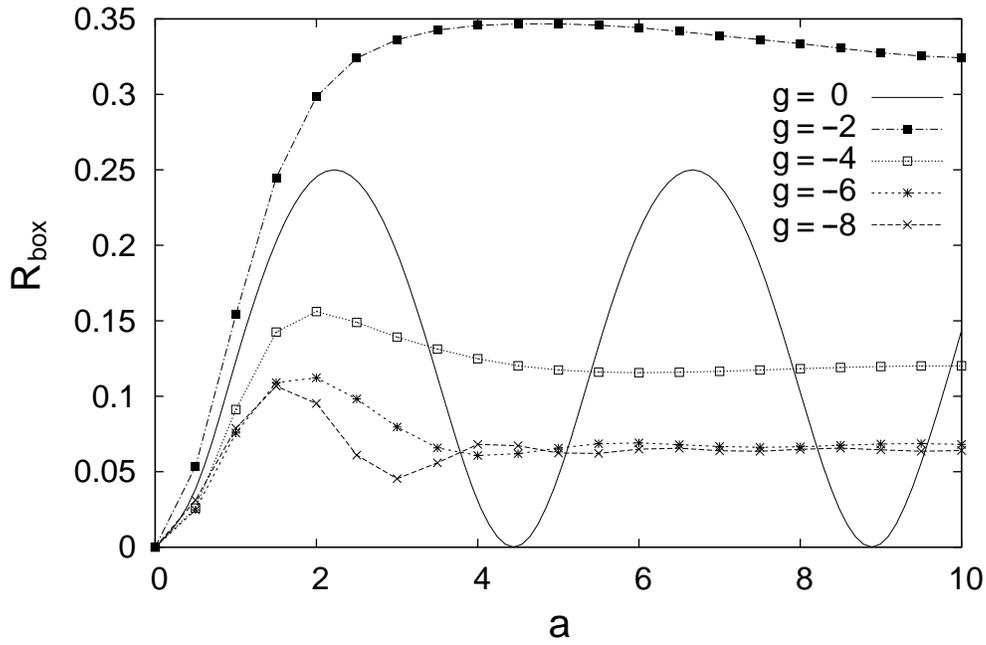}
\end{center}
\caption{Reflectance $R_{\mathrm{box}}$ from the box-type potential
 (\ref{box}) for various values of $g$. The initial condition is the
 Gaussian-type wave packet (\ref{Gaussian}) with $x_{0}=5$ and
 $v=\sqrt{1.5}$, except for $g=0$. The curve for $g = 0$ corresponds to
the linear case given by eq.~(\ref{rs}).} \label{f4} 
\end{figure}

\begin{figure}
\begin{center}
\includegraphics{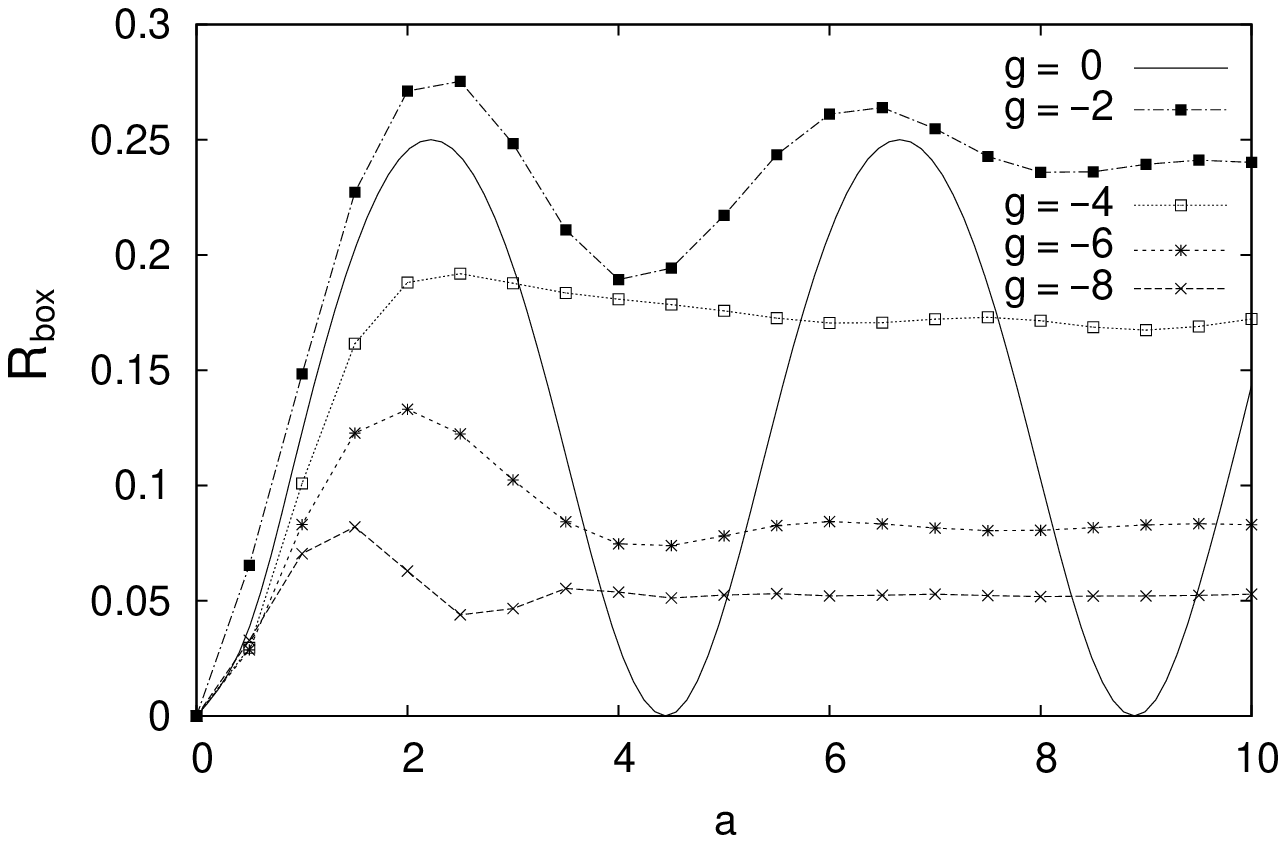}
\end{center}
\caption{Reflectance $R_{\mathrm{box}}$ from the box-type potential
 (\ref{box}) for various values of $g$. The initial condition is the
 Gaussian-type wave packet (\ref{Gaussian}) with $x_{0}=100$ and
 $v=\sqrt{1.5}$, except for $g=0$. The curve for $g = 0$ corresponds to
the linear case given by eq.~(\ref{rs}).} \label{f5}
\end{figure}

\begin{figure}
\begin{center}
\includegraphics{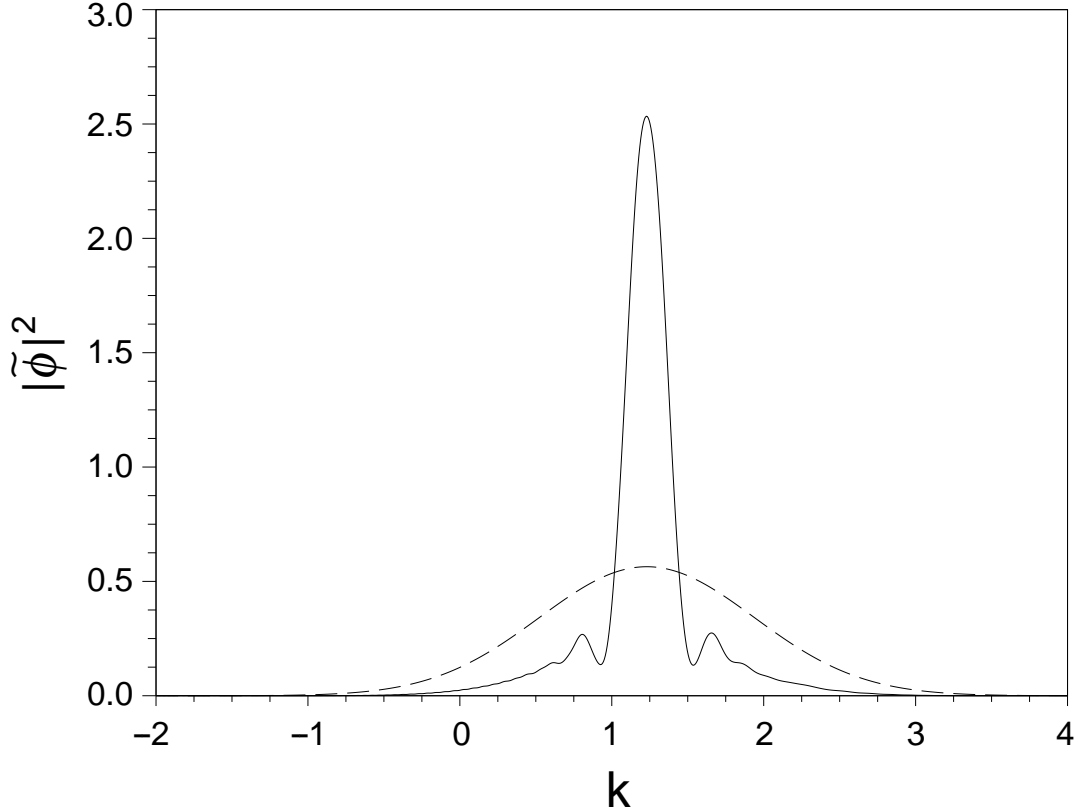}
\end{center}
\caption{Solid line shows $|\tilde{\phi}|^2$ interacting with the box-type potential with width $a=0.5$. The starting point is $x_0=100$ and $g=-2$. This is a snapshot taken at $t=26$. The dashed line is for $t=0$.} \label{nf5} 
\end{figure}

\begin{figure}
\begin{center}
\includegraphics{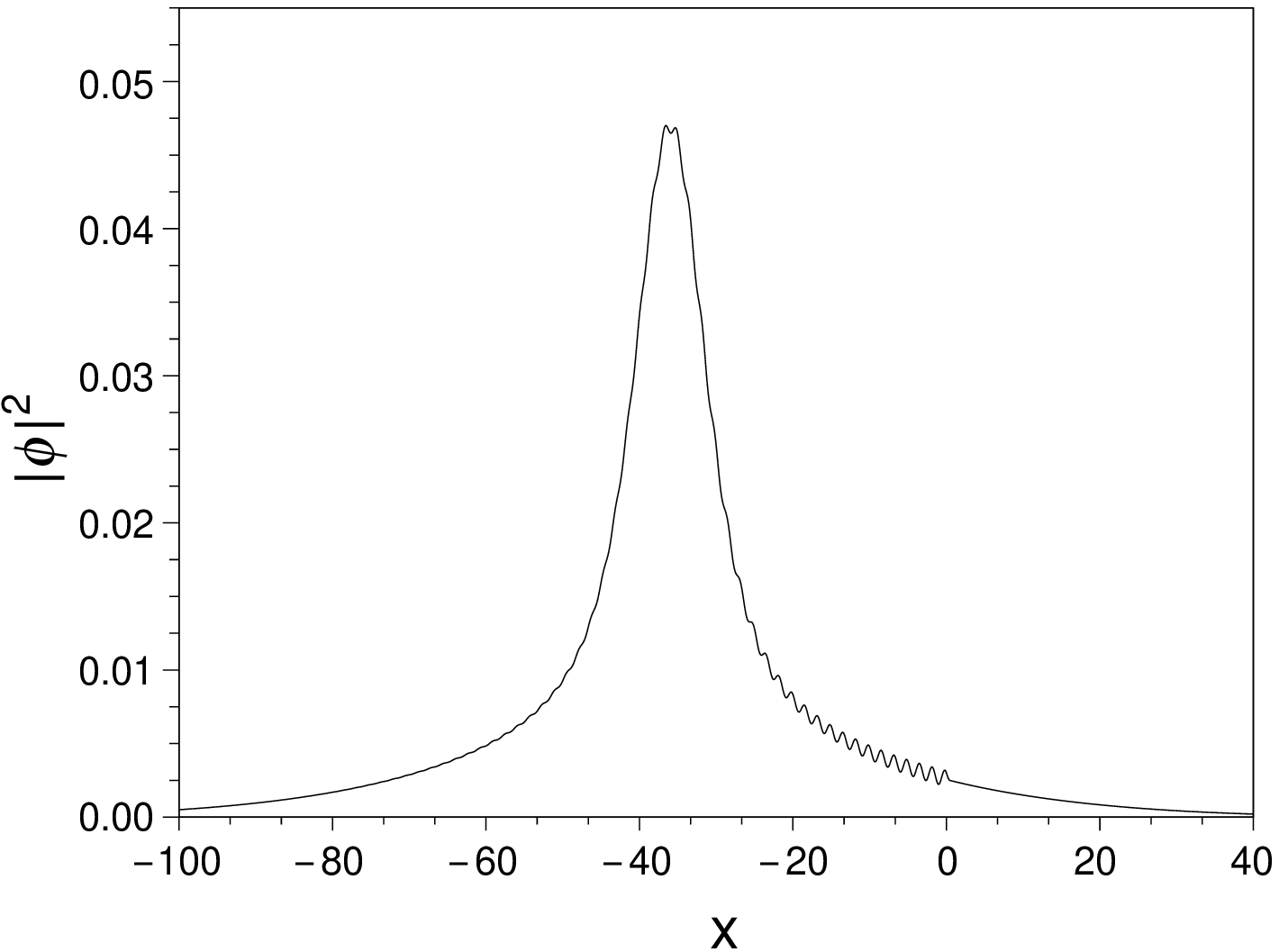}
\end{center}
\caption{Wave profile $|\phi|^2$ colliding at the box-type potential. The parameters are the same  as the ones used in the case ofFig.~\ref{nf5}.} \label{nf6}
\end{figure}
We move on to the main issue: the influence of the wave packet and nonlinear effects on the potential scattering problems. In this subsection, we
consider a box-type (repulsive) potential (\ref{box}).

When we consider a stationary problem with $g=0$, this is a textbook example of quantum mechanics where the analytic expression for the reflectance is
obtained. The expression reads 
\begin{equation}
R_{\mathrm{s}} = \left[1+\frac{4v^2(v^2-1)}
{\mathrm{sin}^2(a\sqrt{v^2-1})}\right]^{-1}\label{rs}, 
\end{equation}
where $v$ is twice the wave number of the incident plane
wave. When $\mathrm{sin}(a\sqrt{v^2-1})=0$, $R_{\mathrm{s}}$ becomes 0
and perfect transmission is realized. This is a kind of resonance
scattering. 

Hereafter, we set $x_{0}$ in eq.~(\ref{Gaussian}) to be 5
or 100 throughout this paper. The parameter $v$ is also fixed at
$\sqrt{1.5}$. 

 The dependences of $R_{\mathrm{box}}$ on $g$ and $a$ are shown in
 Figs.~\ref{f4} and \ref{f5}. The former is for $x_{0}=5$ and the
 latter, $x_{0}=100$. The curve for $g=0$ corresponds to the linear case given by eq.~(\ref{rs}), the reflectance calculated from the stationery Sch\"odinger equation. As mentioned above, the quantity $R_{\mathrm{s}}$ becomes 0 values as $a$ increases. This is due to resonance and is also expected to occur
 periodically as the value of $a$ grows larger. 

The behavior of $R_{\mathrm{box}}$ given by TDGPE (\ref{TDGPE}) is
drastically different. Firstly, maximum values of $R_{\mathrm{box}}$
for each $g$ are totally suppressed for $g<-2$, although
$R_{\mathrm{box}}$ is enhanced for the case of $g=-2$. Secondly, they
never experience the perfect transmission resulting from the wave
packet effect (This is a common feature of wave packet scattering. Such a result is expected even if we set $g=0$, since the measure of the resonant monochromatic plane waves composing the wave packet is zero.), and they seem to be approaching their own constant values asymptotically accompanied with small oscillation as $a$ increases, i.e., the periodic resonance structure is destroyed for $g<-2$ cases. 

This destruction should be regarded as the nonlinear effect because restoration phenomena of the resonance structure are expected if we use linear or weakly interacting wave packets starting from $x_0=100$. In fact, a wavy resonance structure seems to recover for the weakly interacting ($g=-2$) case after long free propagation, as shown in Fig.~\ref{f5}. This can be reasoned as follows: since relatively weak nonlinearity of $g=-2$ cannot prevent the wave packet from diffusing, it spreads and gains sufficient width for the plane wave approximation
to be applied after long propagation. Therefore, the result approaches
the linear case.

To examine the validity of the plane wave approximation, we present Figs.~5 and 6. The former is the profile $|\tilde{\phi}|^2$ in the wave-number space interacting with the box-type potential with width $a=0.5$. The starting point is $x_0=100$ and $g=-2$. This is a snapshot taken at $t=26$. The latter is the wave profile $|\phi|^2$ in the real space colliding at the same box-type potential. The parameters are the same  as the ones used in the case of Fig.~\ref{nf5}. As we have shown in Fig.~5, $|\tilde{\phi}|^2$ for $g=-2$ after propagation is concentrated around the center of the original profile. This means the spectral profile for $g=-2$ still keeps being localized. In contrast, the wave function for $g=-2$ in real space (Fig.~6) broadly spreads over the potential ($a=0.5$). This implies the plane wave approximation is valid. Therefore, for the linear or the sufficiently weakly interacting case, the behavior of $R_{\mathrm{box}}$ for wave packets starting from $x_0=100$ is expected to be reminiscent of the plane waves and to exhibit the restoration phenomena.

However, the restoration never occurs for strongly interacting wave packets ($g<-2$), which means the destruction of the resonance structure is due purely to the nonlinear effect.  

\subsection{Well-type potential}
\begin{figure}
\begin{center}
\includegraphics{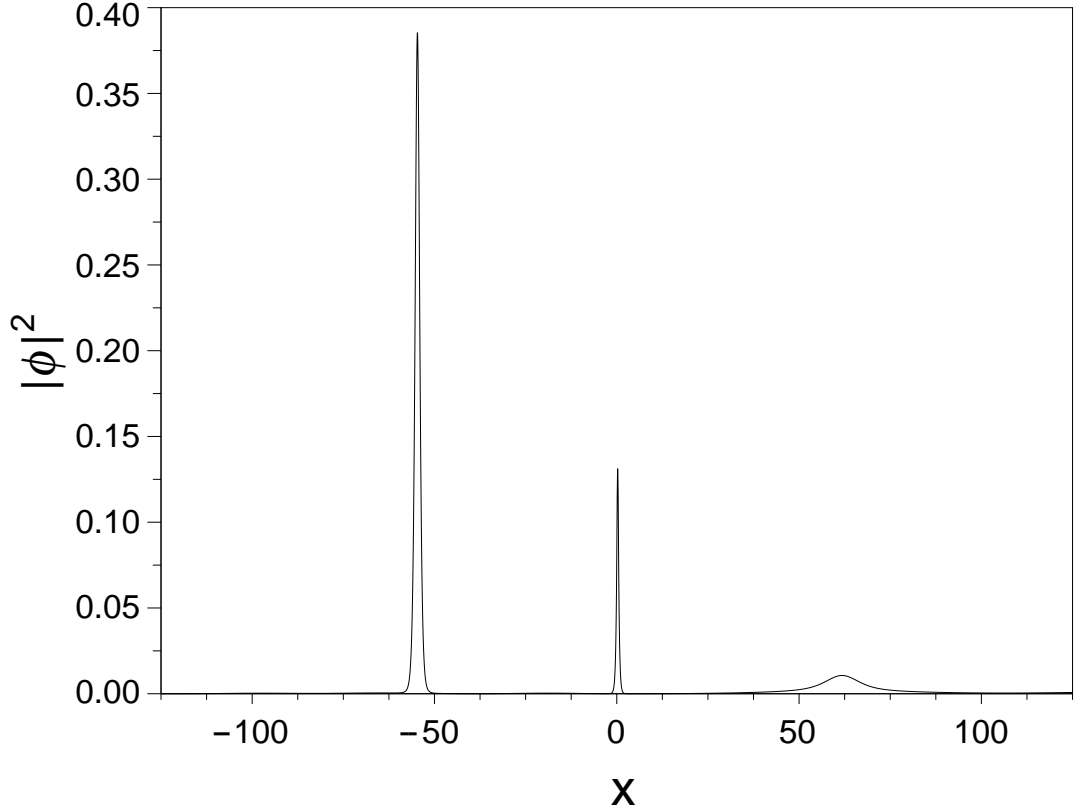}
\end{center}
\caption{Typical wave shape including the portion trapped by
 attractive well-type potential (\ref{well}) with $a=0.5$. The wave
 packet located near the origin is the trapped portion. The initial
 condition is the Gaussian type wave packet (\ref{Gaussian}) with
 $x_{0}=5$, $v=\sqrt{1.5}$ and $g=-8$. This figure shows the snapshot
 taken at $t=30$. $T_{\mathbf{well}}$=0.233. } 
\label{f6}
\end{figure}

\begin{figure}
\begin{center}
\includegraphics{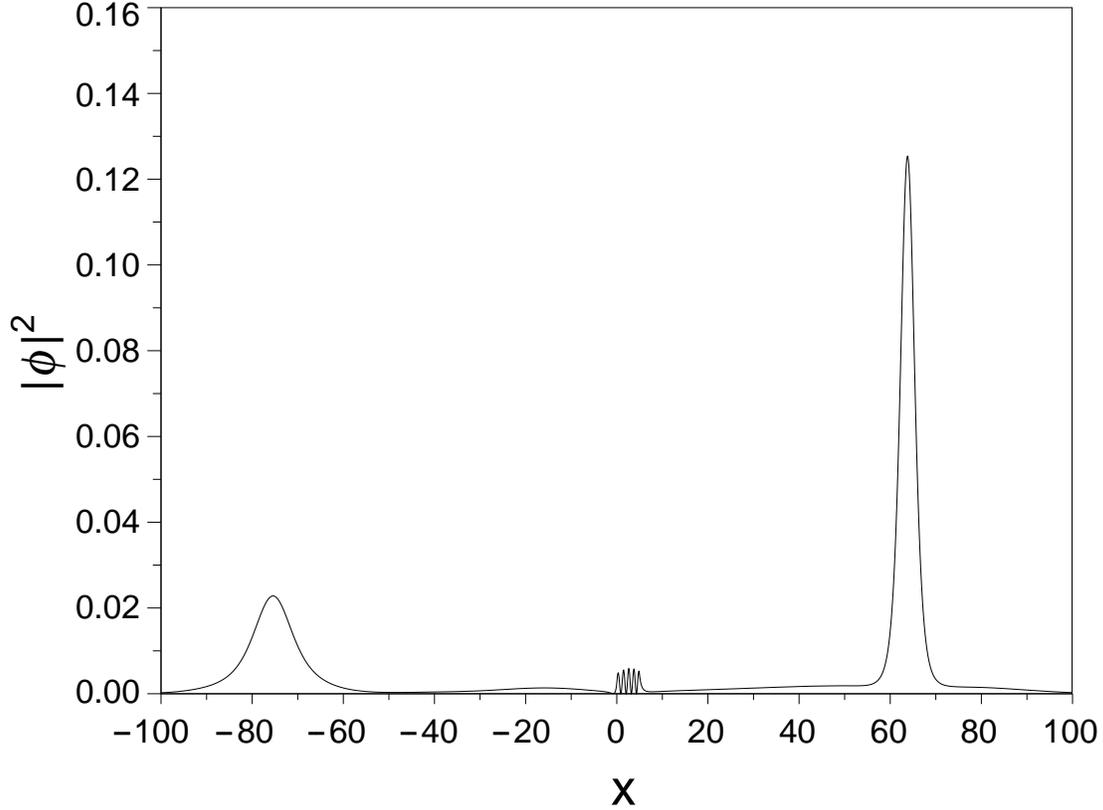}
\end{center}
\caption{Typical wave shape including the portion trapped  by
 attractive well-type potential (\ref{well}) with $a=5$. The trapped
 portion forms a standing-wave-like structure in the potential well. The
 initial condition is Gaussian-type wave packet (\ref{Gaussian})
 with $x_{0}=5$, $v=\sqrt{1.5}$, and $g=-4$. This figure shows the snapshot taken at $t=30$. $T_{\mathbf{well}}$=0.640.} \label{f7}
\end{figure}

\begin{figure}
\begin{center}
\includegraphics{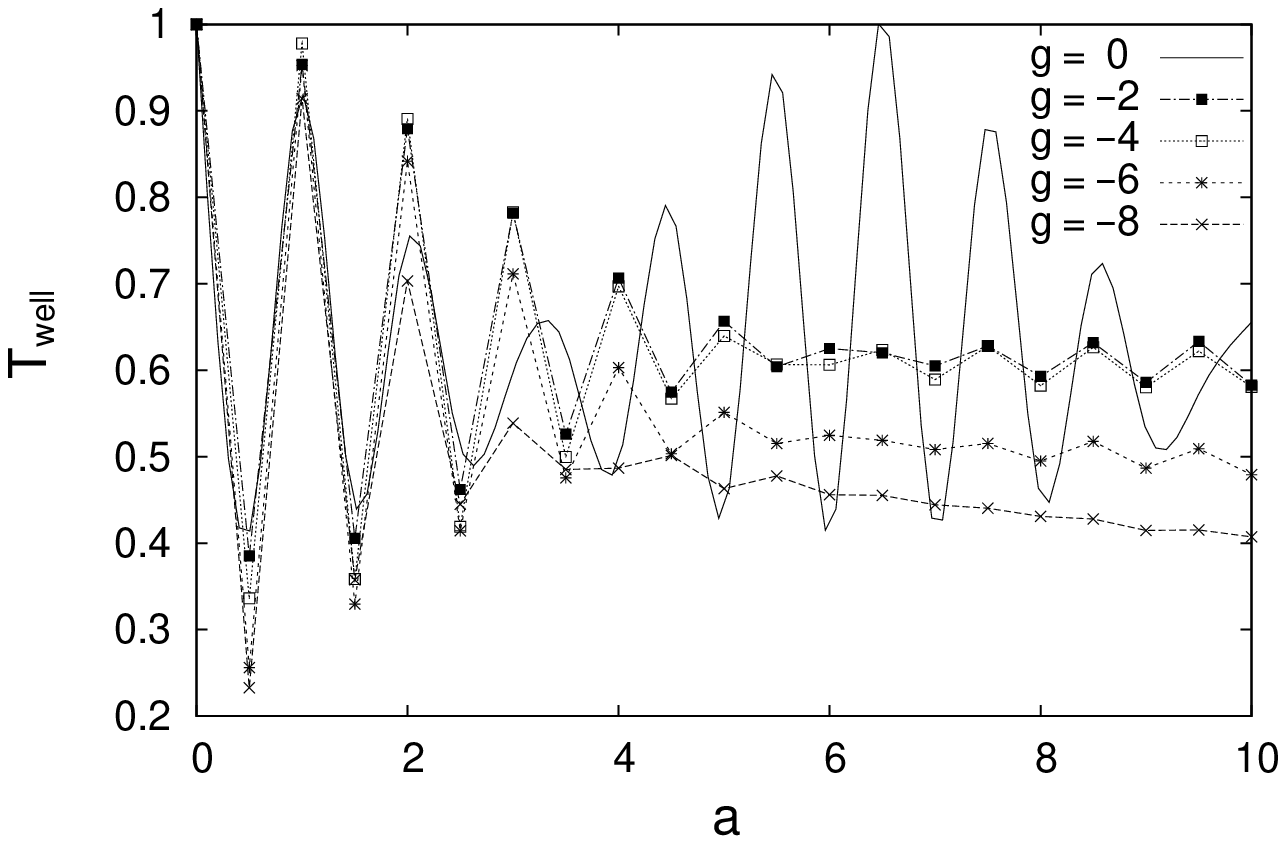}
\end{center}
\caption{Transmittance $T_{\mathrm{well}}$ over the well-type
 potential (\ref{well}) for various values of $g$. The initial
 condition is the Gaussian-type wave packet (\ref{Gaussian}) with
 $x_{0}=5$ and $v=\sqrt{1.5}$. The curve for $g = 0$ corresponds to the linear case given by eq.~(\ref{ts}).} \label{f8}
\end{figure}

\begin{figure}
\begin{center}
\includegraphics{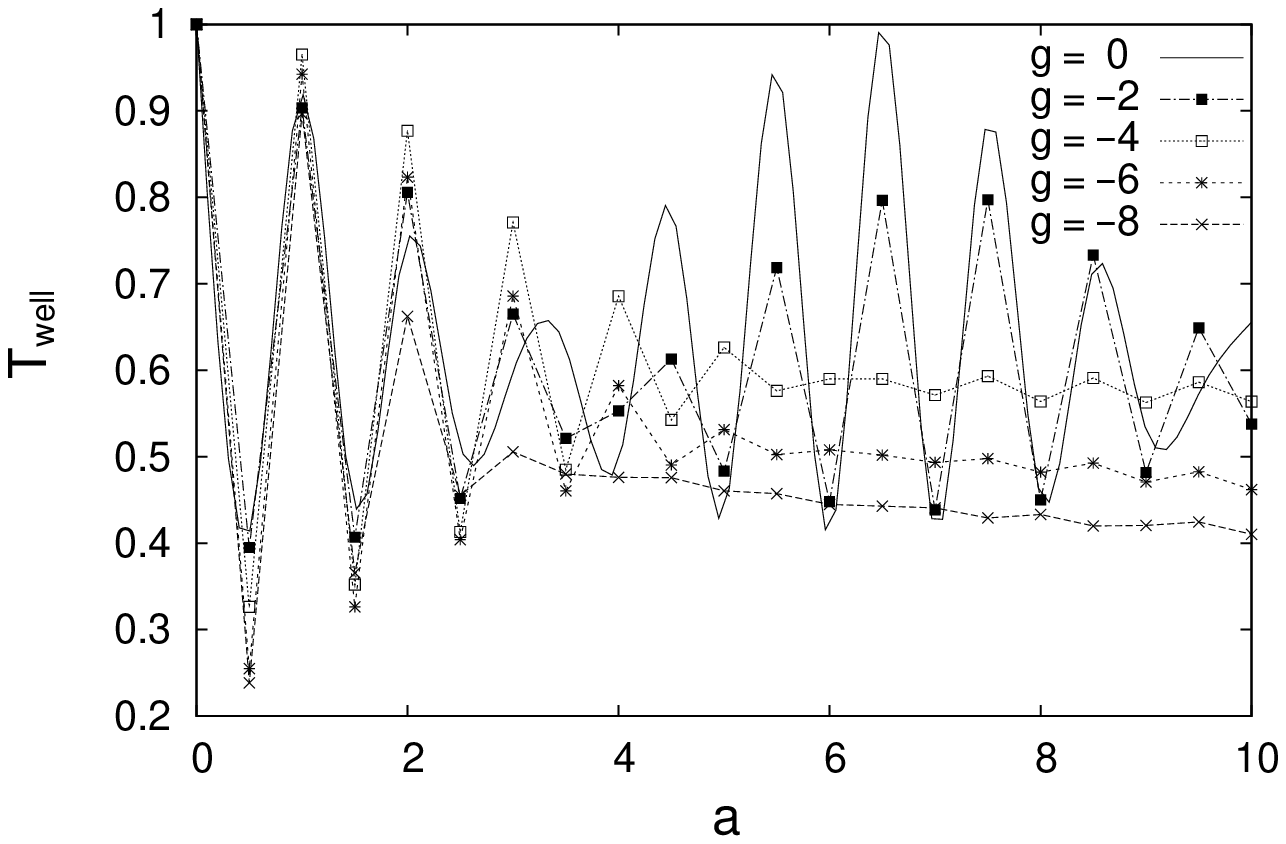}
\end{center}
\caption{Transmittance $T_{\mathrm{well}}$ over the well-type
 potential (\ref{well}) for various values of $g$. The initial
 condition is the Gaussian-type wave packet (\ref{Gaussian}) with
 $x_{0}=100$ and $v=\sqrt{1.5}$. The curve for $g = 0$ corresponds to the linear case given by eq.~(\ref{ts}).} \label{f9}
\end{figure}

\begin{figure}
\begin{center}
\includegraphics{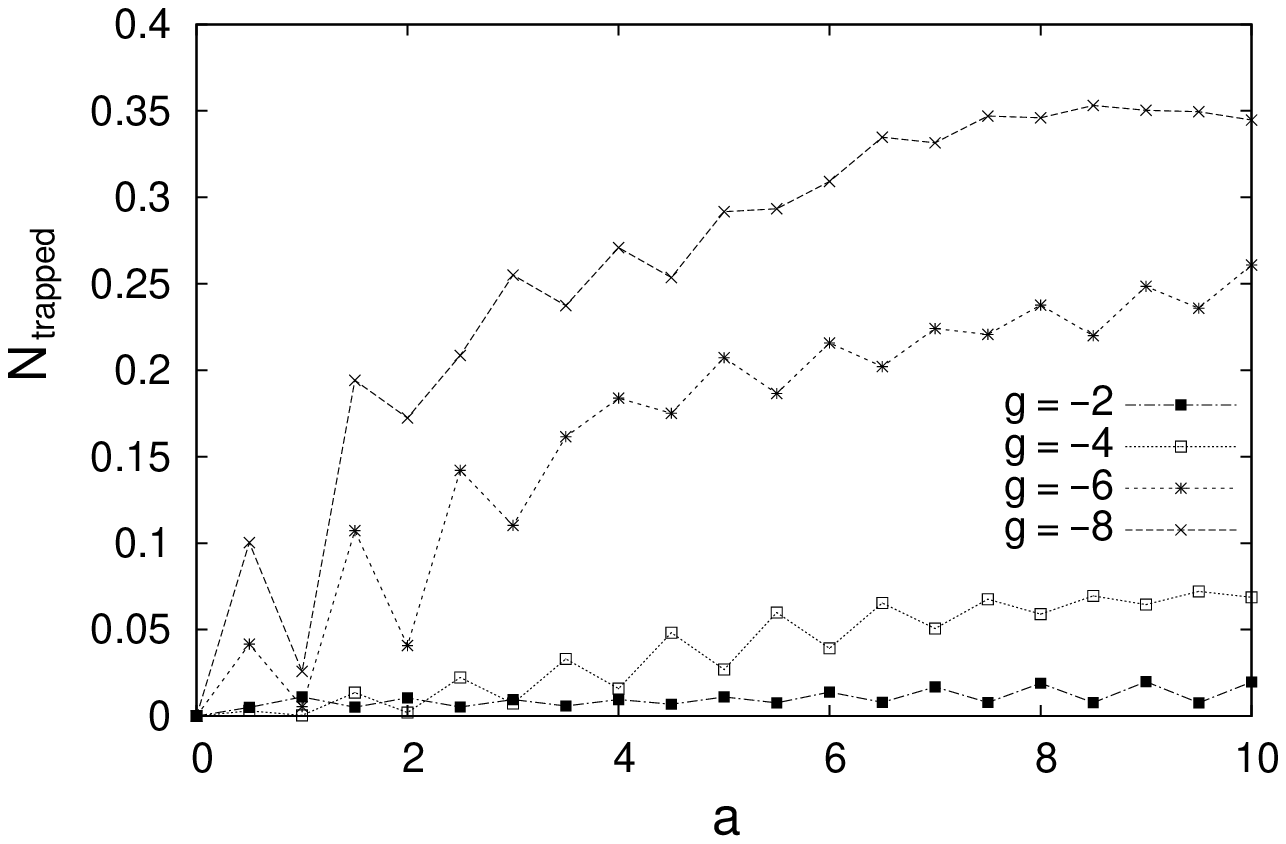}
\end{center}
\caption{Portion trapped $N_{\mathrm{trapped}}$ by the well-type
 potential (\ref{well}) for various values of $g$. The initial
 condition is the Gaussian-type wave packet (\ref{Gaussian}) with
 $x_{0}=5$ and $v=\sqrt{1.5}$.} \label{f10}
\end{figure}

\begin{figure}
\begin{center}
\includegraphics{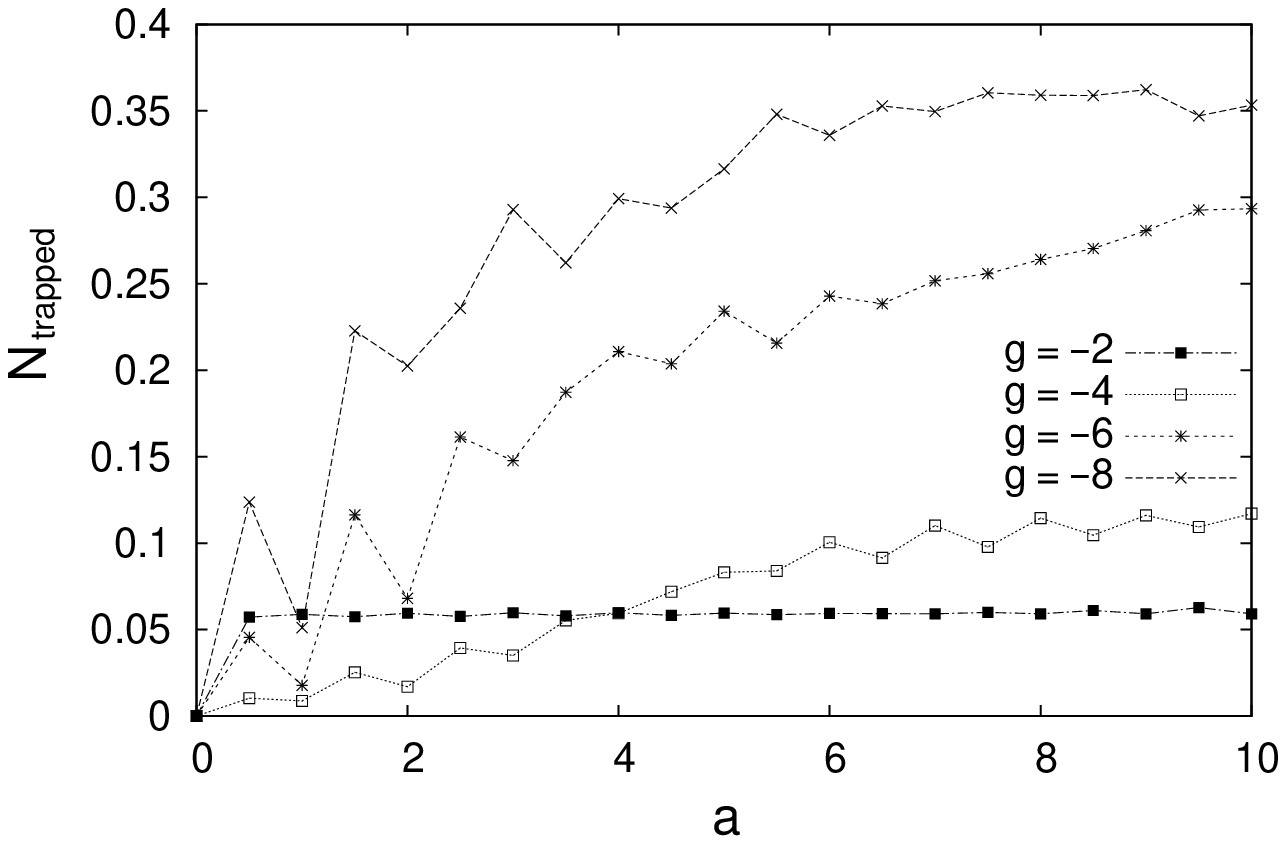}
\end{center}
\caption{Portion trapped$N_{\mathrm{trapped}}$ by the well-type
 potential (\ref{well}) for various values of $g$. The initial
 condition is the Gaussian-type wave packet (\ref{Gaussian}) with
 $x_{0}=100$ and $v=\sqrt{1.5}$.} \label{f11}
\end{figure}

Next, we move on to the well-type (attractive) potential case. The
potential is shown by eq. (\ref{well}). The parameters are the same as those in
the previous case, but the potential depth $V_{0}$ is taken to be
10. For the stationary and linear case, the analytical expression for the
transmittance is given as
\begin{equation}
T_{\mathrm{s}} =
\left[1+\frac{100\,\mathrm{sin}^2(a\sqrt{v^2+10})}{4v^2(v^2+10)}\right]^{-1}\label{ts}. 
\end{equation}
Perfect transmission is realized when $\mathrm{sin}(a\sqrt{v^2+10})=0$.
However, in the nonlinear wave packet dynamics, we require more
careful definition of the reflectance and the transmittance, since a certain amount of wave packets is trapped by attractive potentials. For example, snapshots of the trapped wave profiles are
shown in Figs.~\ref{f6} and \ref{f7}. We can observe a standing-wave-like structure in the latter, and it swings back and forth in the
potential area. Moreover, these trapping phenomena seem to be an intermediate state and the trapped parts continue to gradually emit a part of themselves mainly toward the left. Therefore, the reflectance defined by
eq.~(\ref{rbox}) never converges, even after a very long time. Here, we
employ expression (\ref{rwell}) or (\ref{twell}) instead of eq.~(\ref{rbox}) to evaluate the wave packet and nonlinear effects. The value
of margin $b$ is chosen to be 30 in this work.

The dependences of $T_{\mathrm{well}}$ on $g$ and $a$ are shown in
Figs. \ref{f8} and \ref{f9}. The former is for $x_{0}=5$ and the
latter, $x_{0}=100$. The curve for $g=0$ corresponds to the linear case given by eq.~(\ref{ts}), the transmittance was calculated from
 the stationery Sch\"odinger equation. The quantity $T_{\mathrm{s}}$ becomes unity several times as $a$ increases. This is also due to resonance and also is expected to occur periodically as the values of $a$ grows larger. 

The main features resemble those of the box-type
potential case. Firstly, the maximum values of $T_{\mathrm{well}}$ for
each $g$ are totally suppressed for $g<-2$, although
$T_{\mathrm{well}}$ is enhanced for the case of $g=-2$. Secondly, they
never experience the perfect transmission resulting from the
wave packet effect, and periodic resonance structure is destroyed for strongly self-focusing wave packets ($g<-2$). Thirdly, the wavy resonance structure seems
to recover in the $g=-2$ case after long free propagation
(Fig.~\ref{f9}). The reason for this restoration seems to be the
same as that in the box-type potential case. 

Here, we evaluate the amount of trapped portions by subtracting the sum
of reflectance (\ref{rwell}) and transmittance (\ref{twell})
from unity, i.e.,
\begin{equation}
N_{\mathrm{trapped}} = \lim_{t\rightarrow\infty}
 \int_{-30}^{a+30}|\phi|^2\mathrm{d}x. 
\end{equation}
As mentioned before, these values are nothing more than estimates from
the values at $t=80$. Figures \ref{f10} and \ref{f11} show the
dependences of $N_{\mathrm{trapped}}$ on $g$ and $a$. They basically
show that the amount of trapped portion rises as $|g|$ and $a$
increase, except for the relation between cases of $g=-2$ and $-4$ in the
small $a$ region. We attribute this phenomenon to a) stronger attractive interaction that contributes more to make the value of eq.~(\ref{local}) negative. In the next section, we explain that the negative value of eq.~(\ref{local}) is relevant to the trapping phenomenoa; b) wide potential width, where as it grows, the trapping capacity increases. Finally, we mention that the wavy structures seem to be accompanied with the resonances. 

\section{Discussions}
We studied the scattering problems of nonlinear wave packets as described in
the previous section. One of the noteworthing properties of these problems
is that the final reflectance or transmittance is a function of not only $v$ but also the initial position of the wave packet. In \S2, we showed strong modulation of the Fourier spectrum of a wave
packet due to the nonlinearity which is shown in Fig.~\ref{f2}. The
shape of the Fourier spectrum deforms and oscillates moment by moment 
during propagation. Therefore, the Fourier spectrum at the moment when the wave
packet arrives at the potential area depends on the parameter $x_{0}$,
i.e., the distance between the starting position of the wave packet
and the potential. The shape of the Fourier spectrum evidently affects the reflectance or the transmittance. This is the reason why the final results depend on $x_{0}$. On the contrary, for the linear case, the initial Fourier
spectrum is conserved under free propagation, and the role of $x_{0}$
is not important.

The initial position of the wave packet also affects the result of the scattering problem through the alteration of the incident kinetic energy. 
The kinetic energy $K$ is defined as 
\begin{equation}
K=\int|\phi_{x}|^2\mathrm{d}x\label{kinetic},
\end{equation}
and the self-interaction energy as
\begin{equation}
I=\frac{1}{2}g\int|\phi|^4\mathrm{d}x\label{self}.
\end{equation}
The breathing wave packet is always exchanging its kinetic and
self-interaction energy even during free propagation. As we can see
from eq. (\ref{kinetic}), when the wave packet becomes steeper, the
kinetic energy increases. Since total energy $E=K+I$ is a conserved
quantity in the case of free propagation, the negative self-interaction energy decreases to compensate
the increase of kinetic energy. Therefore, the incident kinetic energy
is also a function of the initial position of the wave packet
$x_{0}$. This incident kinetic energy directly fixes the wave number
at the incident wave packet and becomes one of the most significant factors of the scattering problem. 

From the above considerations, any argument on potential scattering
problems of a nonlinear wave packet requires the consideration of the
initial position of the wave packet $x_{0}$, except the soliton
initial condition. In this work, we fixed $x_{0}$ to be 5 or 100.  
The reflectance and transmittance might be altered for a different
choice of $x_0$, while our main arguments are retained, i.e., the decay of
the resonance structure and the existence of a trapped portion for well-type
potentials might be observed.  

In the previous section, we showed the trapping effect for
self-focusing wave packets by an attractive potential. This phenomenon
is interpreted as a purely nonlinear one. For linear quantum mechanics, this kind of phenomenon never occurs owing to the prohibition of energy level crossing between scattering and bound states. This is because time evolution of a wave packet from an initial state to a final one is fully described by a superposition of elements in the complete set of scattering state eigenfunctions, i.e., 
\begin{equation}
|\phi(0) \rangle = \int c_{\lambda}|E_{\lambda}\rangle
 \mathrm{d} \lambda \Rightarrow |\phi(t)\rangle = \int c_{\lambda}
 \mathrm{e}^{-\mathrm{i}E_{\lambda}t}|E_{\lambda}\rangle
 \mathrm{d}\lambda, 
\end{equation}
where $\lambda$ represents continuous energy eigenvalues and the integral
is taken over all the scattering eigenstates. 
For nonlinear wave packets, however, the distribution of the conserved 
energy is always changing, as argued above, and a contribution from potential energy,
\begin{equation}
V=\int_{0}^{a}V(x)|\phi|^2\mathrm{d}x,
\end{equation}
appears in the potential area. As seen from Figs.~\ref{f6} and \ref{f7}, the trapped parts of the wave packets are well separated from other parts, and hence it makes sense for us to define 
\begin{equation}
(K+I+V)_{\mathrm{local}}=\int_{0}^{a}(|\phi_{x}|^2+V(x)|\phi|^2+\frac{1}{2}g|\phi|^4)\mathrm{d}x\label{local},
\end{equation}
locally. If we calculate ea.~(\ref{local}), we actually get negative values. This trapped state can be considered a ``dynamical bound state'', since a part of the wave function is trapped and continues to oscillate dynamically, staying at around a potential having negative energy.  
 
It is worth mentioning that for the case of the box-type potentials,
the reflectance of strongly self-focusing wave packets is observed
to approach constant values for larger potential width $a$, and its 
dependence on parameter $a$ disappears (Figs.~\ref{f4} and
\ref{f5}). Because the norms of the self-focusing wave packets have 
finite values only in the narrow limited areas between potential ends
and do not have sufficient extent to cover the whole potential area, the wave packets become insensitive to the opposite far end of the potential. In general, the resonance is a result of interference
between the forward-propagating wave and the backward propagating
one. However, the squeezed wave packets merely have a backward
portion since they negligibly interact with the other far end of the potential.
Such wave packets undergo the effective potential for large $a$,
\begin{equation}
V_{\mathrm{box-eff}}=\theta(x).
\end{equation}
For well type potentials, the squeezed wave packets first fall off the cliff of the potential,
\begin{equation}
V_{\mathrm{well-eff-1}}=-V_{0}\theta(x).
\end{equation}
Then, they encounter the other side of the potential wall, and they effectively face 
\begin{equation}
V_{\mathrm{well-eff-2}}=V_{0}\theta(x-a).
\end{equation} 
The major part of the transmittance shown in Figs.~\ref{f8} and
\ref{f9} can be considered the remainders after we subtract the
reflectance shown in Figs.~\ref{f4} and \ref{f5} from unity. The
reflected portion by the potential wall repeats reflection in the
valley of the potential and is considered to constitute the dynamical
bound states. 

\section{Summary}
We numerically studied free propagation of wave
packets governed by the TDGPE for various values of coupling constants
$g$. The initial condition was taken to be the Gaussian form, which is
different from the soliton solution. For the strongly self-interacting
wave packets, diffusion in real space was suppressed and they exhibited
breather-like behaviors. In wave-number space, the breathing motion was
also observed, and a notched structure grew on the surface of the wave
packet. 

We also numerically investigated the potential scattering problems
under the same developing equation and initial conditions. The
potential forms were chosen to be the box- or the well-type. We obtained the reflectance $R_{\mathrm{box}}$ and the transmittance
$T_{\mathrm{well}}$ for different values of coupling $g$ and width of the potential $a$, and we compared them with the
predictions made using stationary Sch\"odinger equations. We found that 
the reflectance $R_{\mathrm{box}}$ or the transmittance $T_{\mathrm{well}}$
is a function of the initial position of the wave packet $x_0$. Other roles of
nonlinearity are rather complicated, i.e., it sometimes enhances
$R_{\mathrm{box}}$ or $T_{\mathrm{well}}$, but sometimes the opposite. 
However, there is a tendency that large $|g|$ decreases both
$R_{\mathrm{box}}$ and $T_{\mathrm{well}}$. For larger values of $|g|$
and $a$, $R_{\mathrm{box}}$ and $T_{\mathrm{well}}$ approach constant
values and do not depend on $a$. 

We also observed the dynamically trapped portion of the wave
packet. We estimated its amount, $N_{\mathrm{trapped}}$, by changing 
$g$ and $a$ and found that $N_{\mathrm{trapped}}$ is an increasing
function of $g$ and $a$ except in the small $g$ and $a$ region. Whether this trapping effect is a perpetual or merely transitional one is not obvious and should be deteremined in future work. 
 
Finally, we make some remarks on the possibility of real
experiments. The control of external environments is relatively easy
in the BEC systems where we can confine condensate particles along
a quasi-rectilinear line by tightening the laser beam trap. In addition, we
can freely change the coupling constants by applying the
Feshbach resonance technique\cite{Courteille}. Soliton-like pulses of
BEC have already been created\cite{Khaykovich}. If controllable local
potential are realized, the possibility of observing and confirming our
results by a real experiment is promising. 

\section{Acknowledgment}
The authors would like to express their sincere gratitude for
Professor Miki Wadati for his valuable comments and continuous
encouragement. They are also grateful to Professor Yoshiya Yamanaka
and Professor Ichiro Ohba of Waseda University for their warm
advice. One of the authors, H.F., thanks for Utsunomiya University for offering wonderful working spaces and opportunities for fruitful discussion.

\end{document}